\documentclass[aps,prb,preprint,groupedaddress]{revtex4}
\usepackage[dvips]{graphicx}
\usepackage{epsfig}

\begin{document}
\title{Entanglement as Measure of Electron-Electron 
Correlation in Quantum Chemistry Calculations} 
\author{Zhen Huang and Sabre Kais\footnote{
Corresponding author:  kais@purdue.edu}\footnote{
Accepted for publications in Chemical Physics Letters, 2005}}
\affiliation{Department of Chemistry, Purdue University, West Lafayette, IN 47907} 

\begin{abstract}
\noindent

In quantum chemistry calculations, the correlation energy
is defined as the difference between the 
Hartree-Fock limit energy and the exact solution of the nonrelativistic Schr$\ddot{o}$dinger 
equation. With this definition,
the electron correlation effects are not directly observable.
In this report, we show that the entanglement can be used as an alternative
measure of the electron correlation in quantum chemistry calculations.
Entanglement is directly observable and it is one
of the most striking properties of quantum mechanics. As an example we calculate 
the entanglement for 
He atom and  H$_2$ molecule with different basis sets.

\end{abstract}

\maketitle
\clearpage

The Hartree-Fock self-consistent
field  approximation, which
is based on the idea that we can approximately
describe an interacting fermion
system in terms of an effective single-particle model,
remains  the starting point and the major approach
for quantitative electronic structure calculations.
In quantum chemistry calculations, the correlation energy
is defined as the energy error
of the Hartree-Fock wave function, i.e., the difference between the 
Hartree-Fock limit energy and the exact solution of the nonrelativistic Schr$\ddot{o}$dinger 
equation\cite{lowdin}. There also exists other measures of electron correlation in the literature
such as the statistical correlation coefficients\cite{kutzelnigg} and more recently
the  Shannon entropy as a measure of the correlation strength\cite{ziesche,qicun}.
Electron correlations have a strong influence on many atomic, molecular\cite{wilson}, and solid properties\cite{march}. Recovering the correlation energy for large systems
remains one of the most challenging problems in quantum chemistry.

The concept of electron correlation as defined in quantum chemistry calculations
is useful but not directly observable,
i.e., there is no operator
in quantum mechanics that its measurement gives the correlation energy. 
In this letter, we propose
to use the entanglement as a measure of the electron correlation.
 Entanglement is directly observable and it is one
of the most striking properties of quantum mechanics. 

It was nearly 70 years ago when Schr$\ddot{o}$dinger gave the name "entanglement" to 
a correlation of quantum nature. He stated that for an entangled state "the
best possible knowledge of the whole does not include the best possible knowledge of 
its parts"\cite{schr}. Over the decades the meaning of the word "entanglement" has 
changed its flavor and our view of the nature of entanglement may continue to be 
modified\cite{burp}.
Entanglement is a quantum mechanical property that describes a correlation 
between quantum mechanical 
systems  that has no classical analog\cite{diviccezo,entg1,Nielsen,gruska}.
A pure state of a pair of quantum systems is called entangled if 
it is unfactoriazable,  as for example,
the singlet state of two spin-$\frac{1}{2}$ particles, 
$\frac{1}{\sqrt{2}} (|\uparrow \downarrow \rangle - |\downarrow \uparrow \rangle)$.
A mixed state is entangled if it can not be represented as a mixture of factorizable
pure states\cite{vedral,hill,Wootters98}. 
Since the seminal work of Einstein, Podolsky, and Rosen\cite{ebr} there 
has been a quest for generating entanglement between quantum 
particles\cite{entg1,blatt}.  Investigation of quantum entanglement is 
currently a very active area
and has been studied intensely
due to its potential applications in quantum communications and 
information processing\cite{entg1} such as quantum teleportation\cite{entg2,dik},
superdense coding\cite{entg3}, quantum key distribution\cite{entg4}, 
telecoloning, and decohernce in quantum computers\cite{div,whaley,omar}.


In order to employ the entanglement as an alternative method of measuring the electron correlation, we present the scheme to quantify  entanglement based on the entanglement measure for two-particle 
systems\cite{Zanardi02,Gittings02}.
We will obtain a general approach 
to quantify the 
entanglement between different spin-orbitals of atomic and molecular systems. 

For two electron system  in $2m$-dimensional spin-space orbital 
with $c_a$ and $c_a^{\dag}$ denote the fermionic annihilation and creation 
operators of single-particle
states and $|0>$ represents the vacuum state, a pure
 two-electron state $|\Phi>$ can be written as 

\begin{equation}\label{OmegaMatrix}
|\Phi> = \sum_{a,b \in \{1,2,3,4,...2m\}}\omega_{a,b}c_a^{\dag}c_b^{\dag}|0>
\end{equation}
where $a, b$ run over the orthonormlized single particle states, and Pauli 
exclusion requires that the $2m \times 2m$ expansion coefficient  matrix  $\omega$ is 
antisymmetric: $\omega_{a,b} = -\omega_{b,a}$, and $\omega_{i,i}=0$ .\\ 

In the occupation number representation $(n_1 \uparrow, n_1 \downarrow,
 n_2 \uparrow, n_2 \downarrow,...,n_m \uparrow, n_m \downarrow)$, 
 where~$\uparrow$~and~$\downarrow$~mean $\alpha$~and~$\beta$ electrons respectively,
 the subscripts denote the spatial orbital index and $m$ is the total spatial orbital number. 
 By tracing out all other spatial orbitals except $n_1$, we can 
 get a ~$(4 \times 4)$~ reduced density matrix for the spatial orbital $n_1$ 
\begin{equation}
\rho_{n_1} = Tr_{n_1}|\Phi><\Phi|=
\left (\begin{array}{cccc}
\rho_{n_1,0} &0 &0 &0\\
0 & 4\sum_{i=1}^{m-1}|\omega_{2,2i+1}|^2 & 0 & 0\\
0 & 0 & 4\sum_{i=2}^m|\omega_{1,2i}|^2 & 0\\
0 & 0 & 0 & \rho_{n_1,2} 
\end{array}\right ),
\end{equation}
where $\rho_{n_1,0}$  denotes an  $''empty~ orbital''$~,
\begin{equation}
\rho_{n_1,0} = 4\sum_{i=1}^{m-1} \sum_{j=1}^{m-1} |\omega_{2i+1,2j+2}|^2
\end{equation}  
and~ $\rho_{n1,2}$ denotes $''two~electron~occupied~orbital''$,
\begin{equation}
\rho_{n1,2} = 4|\omega_{1,2}|^2.
\end{equation}
The $''one~electron~occupied~ orbital''$ , 
in $(\uparrow,\downarrow)$ basis set is given by
\begin{equation}
\rho_{n_1,1}=
\left (\begin{array}{cc}
4\sum_{i=1}^{m-1}|\omega_{2,2i+1}|^2 & 0\\
0 & 4\sum_{i=2}^m |\omega_{1,2i}|^2
\end{array}\right ),
\end{equation}

The matrix elements of $\omega$ can be calculated from the 
expansion coefficient of the ab initio Configure Interaction (CI) method.
The CI wave function with single and double excitation can be written as

\begin{equation}\label{CIExpansion}
|\Phi>=c_0|\Psi_0> + \sum_{ar} c_a^r |\Psi_a^r> + 
\sum_{a<b, r<s} c_{a,b}^{r,s} |\Psi_{a,b}^{r,s}>,
\end{equation}
where $|\Psi_0>$ is the ground state Hartree-Fock wave function, 
$c_a^r$ is the coefficient for single excitation from orbital $a$ to $r$, 
and $c_{a,b}^{r,s}$ is the double  excitation from orbital 
$a$ and $b$ to $r$ and $s$. Now the matrix elements of $\omega$ can be
written in terms of the CI expansion  coefficients
\begin{equation}
\omega_{1,2} = \frac{c_0}{2}, \; \; 
\omega_{2,2i+1} = \frac{c_1^{2i+1}}{2}, \;\;
\omega_{1,2i+2} = \frac{c_2^{2i+2}}{2},\;\;
\omega_{2i+1,2j+2} = -\frac{c_{1,2}^{2i+1,2j+2}}{2},
\end{equation}

where $i,j = 1,2...m$. In this general approach, the ground state entanglement  is given by 
von Neumann entropy  of the reduced density matrix $\rho_{n1}$
\begin{equation}
S(\rho_{n_1}) = -Tr(\rho_{n_1} log_2 \rho_{n_1}).
\end{equation}

We are now ready to evaluate the  entanglement for the H$_2$ molecule as a function of $R$
using a direct and simpler approach based on the two-electron density matrix calculated from
the CI wave function with single and double electronic excitations. In the 
occupation number representation, the CI wave function  is given by  

\begin{equation}
|\Phi> = c_0|1100...> + c_1^3|0110...> + c_2^4|1001...> + c_{1,2}^{3,4}|0011...> + ... .
\end{equation}
By tracing out all other orbitals except $1$, we can get the reduced density 
matrix for $(n_1 \uparrow=0,1)$ 
\begin{equation}
\rho_1 = Tr_1|\Phi><\Phi|= \left (\begin{array}{cc}\sum_{i=1}^{m-1}|c_1^{2i+1}|^2 + \sum_{i=1}^{m-1}|c_{1,2}^{2i+1,2i+2}|^2 & 0 \\
0 & |c_0|^2 + \sum_{i=1}^{m-1}|c_2^{2i+2}|^2 
\end{array}\right ).
\end{equation} 
The CI wave function expansion coefficients are calculated with the electronic structure 
package Gaussian\cite{gaussian}.
Thus the entanglement of H$_2$ molecule is readily calculated by the von Neumann entropy
\begin{equation}\label{H2Entropy}
S(\rho_1) = -Tr(\rho_1 log_2 \rho_1).
\end{equation}

Figure (1) shows the  calculated entanglement $S$ for H$_2$ molecule, Eq. (\ref{H2Entropy}), as a function 
of the internuclear distance $R$ using Gaussian basis
set 3-21G\cite{gaussian}. For comparison we included the usual
electron correlation $(E_c=|E^{Exact}\;-\; E^{UHF}|)$ and spin-unrestricted 
Hartree-Fock (UHF) calculations\cite{gaussian} using the 
same basis set in the figure.
At the limit $R=0$, the dot represents 
the electron correlation 
for the He atom, $E_c=0.0149 (a.u.) $ using 3-21G basis set
 compared with the entanglement for
the He atom $S=0.0313$.  With a 
larger basis set, $cc-pV5Z$\cite{mark},  we obtain numerically $E_c=0.0415(a.u.)$ and $S=0.0675$. 
Thus,
qualitatively entanglement and absolute correlation have similar behavior. 
At the united atom limit,
$R \rightarrow 0$,
both have small values, then rise to a maximum value and finally vanishes at 
the separated atom limit, $R \rightarrow \infty$. However, note that for $R > 3\mathrm{\AA }$
the correlation between the two electrons is almost zero but the entanglement is maximal until around
$R \sim 4\mathrm{\AA }$, the entanglement vanishes for $R > 4\mathrm{\AA }$.

To understand the entanglement behavior for H$_2$ molecule using ab initio quantum 
chemistry methods, we calculate the entanglement for a simpler  two-electron model system. This is a model
of two spin-$\frac{1}{2}$ electrons
with an exchange coupling constant $J$ (a.u.) in an effective transverse magnetic field of strength $B$ (a.u.). In order to describe the environment of the electrons in a molecule, we simply
introduce a small effective external magnetic field $B$. 
 The general Hamiltonian for such a system is given by
\begin{equation}\label{XYHamiltonian}
H = -\frac{J}{2}(1+\gamma)\sigma_1^x \otimes \sigma_2^x - \frac{J}{2}(1-\gamma)\sigma_1^y \otimes \sigma_2^y -B \sigma_1^z \otimes I_2 - B I_1 \otimes \sigma_2^z,
\end{equation}  
where $\sigma^a$ are the Pauli matrices(a= x,y,z) and 
$\gamma$ is the degree of anisotropy. For $\gamma=1$ 
 Eq.(\ref{XYHamiltonian}) reduces to the Ising model, whereas for 
 $\gamma=0$  it is the XY model. 

Our two spin problem admits an exact solution, it is simply a $(4 \times 4)$ matrix
 with the following four 
eigenvalues\\
\begin{equation}
  \lambda_1 = -J, \; \lambda_2 = J, \; 
\lambda_3 = -\sqrt{4B^2 + J^2 \gamma^2}, \; \lambda_4 = \sqrt{4B^2 + J^2 \gamma^2}
\end{equation}
and the corresponding eigenvectors\\
\begin{equation}
|\phi_1> = \left( \begin{array}{c}0\\1/\sqrt{2}\\1/\sqrt{2}\\0 \end{array} \right),~
|\phi_2> = \left( \begin{array}{c}0\\-1/\sqrt{2}\\1/\sqrt{2}\\0 \end{array} \right),~
|\phi_3> = \left( \begin{array}{c}\sqrt{\frac{\alpha +2B}{2 \alpha}}\\0\\0\\
\sqrt{\frac{\alpha - 2B}{2 \alpha}}\end{array}\right),~
|\phi_4> = \left( \begin{array}{c}-\sqrt{\frac{\alpha - 2B}{2 \alpha}}\\0\\0\\
\sqrt{\frac{\alpha + 2B}{2 \alpha}}\end{array} \right),
\end{equation}
where~$\alpha = \sqrt{4B^2 + J^2\gamma^2}$. 
In the  basis set $\{|\uparrow \uparrow>,|\uparrow \downarrow>,|\downarrow \uparrow>,
|\downarrow \downarrow>\}$, the eigenvectors can be written as

\begin{equation}
|\phi_1> = \frac{1}{\sqrt{2}}(|\downarrow \uparrow> + |\uparrow \downarrow>),
\end{equation}
\begin{equation}
|\phi_2> = \frac{1}{\sqrt{2}}(|\downarrow \uparrow> -|\uparrow \downarrow> ),
\end{equation}
\begin{equation}
|\phi_3> = \sqrt{\frac{\alpha - 2B}{2\alpha}}|\downarrow \downarrow> +
\sqrt{\frac{\alpha + 2B}{2\alpha}}|\uparrow \uparrow>,
\end{equation}
\begin{equation}
|\phi_4> = \sqrt{\frac{\alpha + 2B}{2\alpha}}|\downarrow \downarrow> 
-\sqrt{\frac{\alpha - 2B}{2\alpha}}|\uparrow \uparrow>.
\end{equation}

Now we confine our interest to the calculation of the 
entanglement between the two electronic spins. 
For simplicity we take $\gamma =1$, Eq. (\ref{XYHamiltonian}) reduces to the Ising model
with the 
ground state energy $\lambda_3$ and the corresponding eigenvector $|\phi_3>$. All
the information needed for quantifying the entanglement in this case is contained in the 
two-electron density matrix.
 
\indent When a biparticle quantum system AB is in a pure state there is 
essentially a unique measure of the entanglement between the subsystems A and B
 given by the von Neumann entropy $S$\cite{peres}. If we denote  $\rho_A$
 the partial trace of $\rho_{AB}$ with respect to 
 subsystem B, $\rho_A = Tr_B(\rho_{AB})$, the entanglement of the state $\rho_{AB}$ is 
 defined as the von Neumann entropy of the reduced density operator
  $\rho_A$, $S(\rho_{AB})\equiv -Tr[\rho_A log_2 \rho_A]$.\\
\indent For our model  system in the ground state $|\phi_3>$, the 
reduced density matrix in the basis set$(\uparrow,\downarrow)$ is given by
\begin{equation}
\rho_A =\left (\begin{array}{cc}
\frac{\alpha + 2h}{2 \alpha}  & 0 \\
0& \frac{\alpha - 2h}{2 \alpha}
\end{array}\right ).
\end{equation}
Thus,  the entanglement is simply given by
\begin{equation}\label{Entanglement}
S = -\frac{1}{2} log_2(\frac{1}{4} -\frac{1}{4+\lambda^2}) + \frac{1}{\sqrt{4+\lambda^2}}log_2 \frac{\sqrt{4+\lambda^2}-2}{\sqrt{4+\lambda^2}+2}
\end{equation}
where $ \lambda= J/B$.\\
\indent The value of $J$, the exchange coupling constant between the spins of the two electrons,
can be calculated as 
half the  energy difference 
between the lowest singlet and triplet states of 
the hydrogen molecule. Herring and 
Flicker have shown {\cite{Herring64}} that $J$ for H$_2$ molecule can be approximated as a 
function of the interatomic distance $R$. In atomic units, the  expression for
large $R$ is given by
\begin{equation}\label{CouplingConstant}
J(R)=-0.821 \; R^{5/2}e^{-2R}+O(R^{2}e^{-2R}). 
\end{equation} 

 Figure (2) shows the  calculated von Neumann Entanglement (S), Eq. (\ref{Entanglement}), 
 as a function of the distance between the two electronic
spins $R$, using $J(R)$ of Eq. (\ref{CouplingConstant}),
 for different values of the magnetic field strength $B$.  At the limit
$R \rightarrow \infty$ the exchange interaction $J$ vanishes
as a result the two electronic spins are up and the wave function is factorazable, i.e. the
 entanglement is zero. At the other limit, when $R=0$ the entanglement is zero for this model because $J=0$. As $R$ increases,  the exchange interaction increases
  leading to increasing
 entanglement between the two electronic spins.
However this increase in the entanglement 
  reaches a maximum limit as shown in the figure.
For large distance, the exchange interaction  decreases exponentially 
 with $R$ and thus the decrease of the entanglement. 
 The figure also shows that the entanglement  increases with decreasing the magnetic field strength. This can be attributed to effectively increasing the exchange 
 interaction.
Thus we get the similar behavior as the entanglement for the H$_2$ molecule as a function 
 of the internuclear distance $R$ , using accurate ab initio methods. Because the Eq. (\ref{CouplingConstant}) is only applicable to the large value of  $R$, it is not surprising to see, in the limit  $R \rightarrow 0$, the entanglement
 converges to the He atom results in ab initio methods but disappears in this simple model.\\
 
Recently a new promising approach is emerging for the realization of quantum chemistry calculations
without wave functions through first order semidefinite programming\cite{david}. The
electronic energies and properties of atoms and molecules are computable simply from
an effective two-electron reduced density matrix $\rho_{(AB)}$. Thus, the electron correlation 
can be directly calculated as effectively with the entanglement between the two electrons,
which is readily calculated as the von Neumann entropy $S=-Tr \rho_A log_2 \rho_A$, where
$\rho_A=Tr_B \rho_{(AB)}$. Utilizing this combined approach, one calculates the electronic energies
and properties of atoms and molecules including correlation without wave functions or 
Hartree-Fock reference systems.

In summary, we presented the entanglement as an alternative measure of the electron 
electron correlation 
in quantum chemistry calculations for atoms and molecules. All the information needed for
quantifying the entanglement is contained in the two-electron density matrix.
 This measure is readily
calculated by evaluating the von Neumann entropy of the one electron reduced density operator.
This definition of correlation has deep roots in quantum theory, observable, and does 
need a reference 
system such as Hartree-Fock calculations. The approach is general and 
can be used for larger atomic 
and molecular systems.

\begin{acknowledgments}

We would like to thank Professor Dudley Herschbach for critical reading of the manuscript.
This work has been supported by the Purdue Research Foundation.
\end{acknowledgments}

\newpage
\begin{figure}
\begin{center}
\includegraphics[width=1.1\textwidth,height=0.7\textheight]{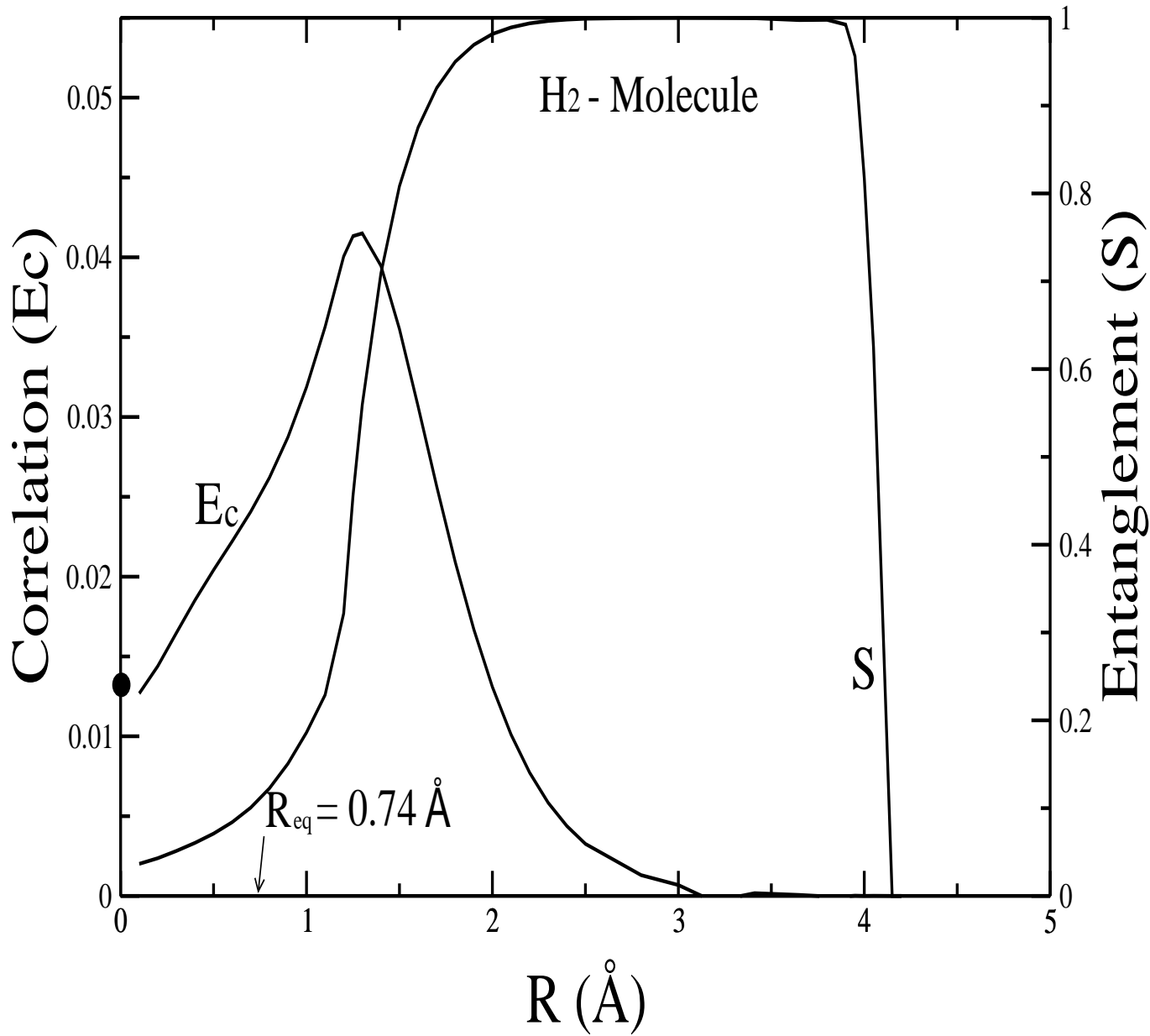}
\end{center}
\caption{Comparison between the absolute value of the electron correlation 
$E_c=|E^{Exact}\;-\; E^{UHF}|$ 
and the von Neumann Entanglement (S) as a function of the internuclear distance $R$ for
the H$_2$ molecule using Gaussian basis set 3-21G. At the limit $R=0$, the dot represents 
the electron correlation 
for the He atom, $E_c=0.0149 (a.u.) $ using 3-21G basis set compared with the entanglement for
He atom $S= 0.0313$. The equilibrium distance using 3-21G basis set is 
$R_{eq}= 0.74\mathrm{\AA }$}.
 \end{figure}

\newpage

\begin{figure}
\begin{center}
\includegraphics[width=1.1\textwidth,height=0.7\textheight]{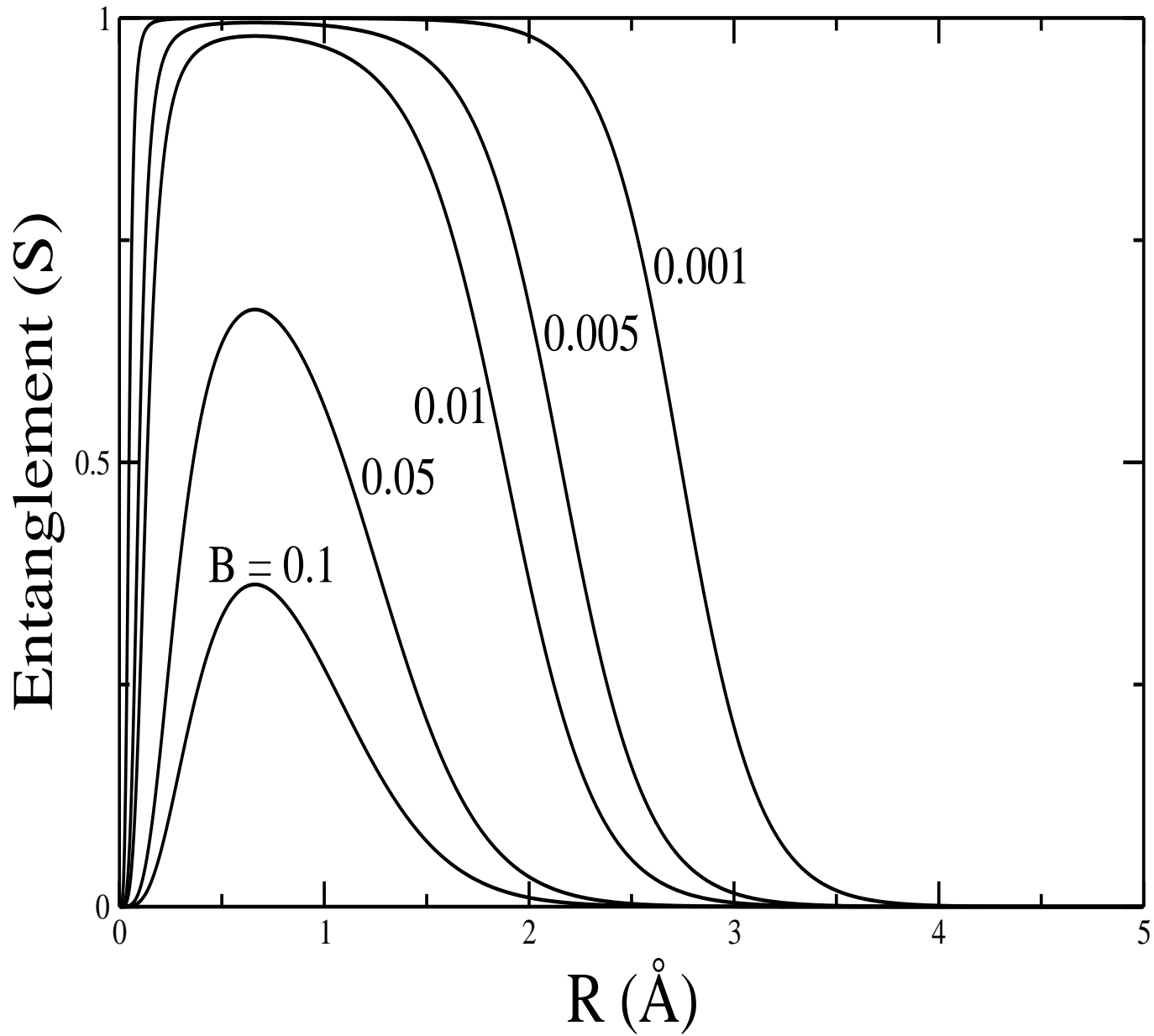}
\end{center}
\caption{von Neumann Entanglement (S) as a function of the distance $R$  between the two
spins  for different values of the magnetic field strength $B$.  }
 \end{figure}

\end{document}